\documentclass[conference]{IEEEtran}
\IEEEoverridecommandlockouts
\renewcommand{\baselinestretch}{1.0} % Change to 1.65 for double spacing
 
\usepackage{amsmath,amsfonts,amssymb}
\usepackage{graphicx}
\usepackage{booktabs}
\usepackage{array} 
\usepackage{caption} 
\usepackage{colortbl} 
\usepackage{xcolor} 
\usepackage{multirow} 

\usepackage[colorlinks=true, allcolors=blue]{hyperref}

\title{Evaluating Deep Learning Models for Breast Cancer Classification: A Comparative Study}

\author{
\IEEEauthorblockN{Sania Eskandari\IEEEauthorrefmark{1},
Ali Eslamian\IEEEauthorrefmark{2},
Nusrat Munia\IEEEauthorrefmark{2},
Amjad Alqarni\IEEEauthorrefmark{2}, and
Qiang Cheng\IEEEauthorrefmark{2}\IEEEauthorrefmark{3}}

\IEEEauthorblockA{\IEEEauthorrefmark{1}Department of Electrical and Computer Engineering, University of Kentucky, Lexington, KY, USA\\
sania.eskandari@uky.edu}

\IEEEauthorblockA{\IEEEauthorrefmark{2}Department of Computer Science, University of Kentucky, Lexington, KY, USA\\
\{ali.eslamian, nusrat.munia, amjad.alqarni\}@uky.edu}

\IEEEauthorblockA{\IEEEauthorrefmark{3}Institute for Biomedical Informatics, University of Kentucky, Lexington, KY, USA\\
qiang.cheng@uky.edu}
}

\begin{document} 
\maketitle

\begin{abstract}
This study evaluates the effectiveness of deep learning models in classifying histopathological images for early and accurate detection of breast cancer. Eight advanced models, including ResNet-50, DenseNet-121, ResNeXt-50, Vision Transformer (ViT), GoogLeNet (Inception v3), EfficientNet, MobileNet, and SqueezeNet, were compared using a dataset of 277,524 image patches. The Vision Transformer (ViT) model, with its attention-based mechanisms, achieved the highest validation accuracy of 94\%, outperforming conventional CNNs. The study demonstrates the potential of advanced machine learning methods to enhance precision and efficiency in breast cancer diagnosis in clinical settings. The code is publicly available at
\url{https://github.com/saniaesk/Breast-Cancer-Classification}
\end{abstract}

% \keywords{ResNet-50, DenseNet-121, ResNeXt-50, Vision Transformer, GoogLeNet, EfficientNet, MobileNet, SqueezeNet.}

\section{INTRODUCTION}

Breast cancer (BC) is a malignant disease originating in breast tissue, often forming detectable tumors identified through imaging methods like mammography \cite{ref1-ara2021malignant}. Early detection is critical but challenging due to the often asymptomatic nature of early-stage BC. Diagnostic methods such as mammography, ultrasound, and biopsy are essential in distinguishing benign from malignant tumors \cite{ref2-le2008gene, arjmand2019breast}. Traditional manual diagnosis is time-consuming and depends on highly skilled pathologists, who can still make errors. To address these challenges, Computer-Aided Diagnostic (CAD) systems have been developed and have significantly aided the diagnostic process \cite{ref3-araujo2017classification}. The success of Convolutional Neural Networks (CNNs) in image classification has encouraged their application in medical imaging, particularly for histopathology image classification \cite{ref4-spanhol2016breast}. CNNs are adept at extracting hierarchical features from images, making them suitable for detecting complex patterns associated with malignant tumors. 

The classification task involves assigning each image or image patch to its appropriate category, such as benign or malignant \cite{ref5-amrane2018breast}. This automated approach aims to improve the accuracy and efficiency of breast cancer detection, potentially leading to earlier diagnosis and better patient outcomes. Studies have shown that Convolutional Neural Networks (CNNs) significantly enhance the classification of breast cancer histology images \cite{ref3-araujo2017classification}. Advances in deep learning, including models like ResNet, DenseNet, and Vision Transformers, have shown promising results in medical image processing by using complex structures, residual connections, and attention mechanisms to improve performance  \cite{ref6-amrane2018breast, posso2023non}. Additionally, other machine learning methods, such as Support Vector Machines (SVMs), have been used for gene selection and cancer categorization, highlighting the versatility of machine learning in oncology \cite{ref2-le2008gene}. Advanced models like EfficientNet and MobileNet have also been adapted for medical applications, balancing precision and computational efficiency \cite{ref7-tan2019efficientnet}. In addition to this, the adaptive vector quantization model effectively clusters high-dimensional medical data \cite{deldadehasl2025dynamic}.

This research aims to identify the most effective methods for classifying histopathology images of breast cancer through a comparative analysis of eight advanced models: ResNet-50, DenseNet-121, ResNeXt-50, Vision Transformer (ViT), GoogLeNet (Inception v3), EfficientNet, MobileNet, and SqueezeNet. The goal is to enhance diagnostic accuracy, leading to better patient outcomes and potentially saving lives.

\section{PROPOSED APPROACH}
This study aims to identify the most accurate model for classifying breast cancer from histopathology images by comparing various modern CNN models and the attention-based Vision Transformer. The selected models include ResNet-50, DenseNet-121, ResNeXt-50, Inception v3, EfficientNet, MobileNet, SqueezeNet, and ViT. Each model's potential effectiveness in medical image analysis will be systematically evaluated to enhance the accuracy and efficiency of breast cancer diagnosis.

\section{RESULTS AND DISCUSSION}

% \noindent\textbf{Dataset}
\subsection{Dataset}

The Breast Histopathology Images dataset comprises histopathology photographs of breast biopsies, specifically focusing on instances of invasive ductal carcinoma (IDC), the most prevalent subtype of breast cancer \cite{ref8-janowczyk2016deep}. Pathologists often concentrate on areas containing IDC when assessing the aggressiveness of a whole mount sample. Therefore, an essential first step in automating this assessment is to accurately identify the specific areas of IDC within an entire slide. The initial dataset consists of 162 whole-mount slide images of breast cancer specimens, scanned at a magnification of 40x. From these slides, a total of 277,524 patches, each measuring 50×50 pixels, were extracted. The patches are categorized based on the presence or absence of IDC, with 198,738 patches labeled as IDC negative and 78,786 patches labeled as IDC positive.

To train and evaluate the model, the dataset was divided into three subsets: training data, validation data, and test data. This partitioning ensures that the model can be trained efficiently, validated during the training phase to optimize hyperparameters, and assessed on a separate dataset to evaluate its generalization ability. Table \ref{table:Dataset Partitioning} summarizes the distribution of samples among these subsets.

\begin{figure}
\centering
\includegraphics[width=\linewidth]{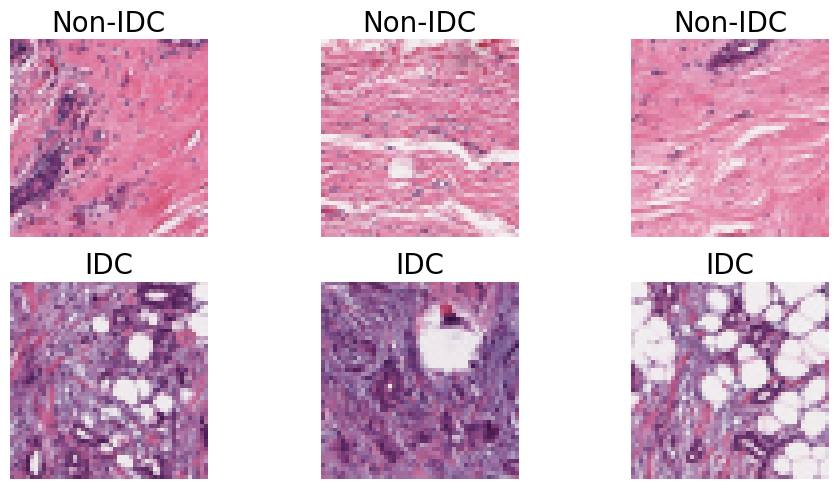}
\caption{Examples of histopathology images. Top row: Non-IDC, Bottom row: IDC. Images sourced from Kaggle \cite{ref16-he2016deep}}\label{fig:Dataset}
\end{figure}

\begin{table}[ht]
    \centering
    \caption{Dataset Partitioning}
    \label{table:Dataset Partitioning}
    \begin{tabular}{lccc}
        \toprule
        Subset & \textbf{IDC Negative} & \textbf{IDC Positive} & \textbf{\textit{Total}} \\
        \midrule
        Training & 138,288 & 55,978 & 194,266 \\
        Validation & 39,748 & 15,760 & 55,508 \\
        Test & 20,702 & 7,048 & 27,750 \\
        \midrule
        Total & 198,738 & 78,786 & 277,524 \\
        \hline
    \end{tabular}
\end{table}

Through careful and systematic preparation and division of the dataset, our goal is to ensure that our deep learning models are trained on a comprehensive and representative collection of data. This will enable them to perform effectively in accurately diagnosing IDC in breast histopathology images.

% \noindent\textbf{Experimental Setup}
\subsection{Experimental Setup}

The experimental system architecture was deployed on a system equipped with an AMD Ryzen Threadripper PRO 5965WX 24-core CPU, 62 GB of RAM, and an NVIDIA RTX A4500 GPU with 20 GB of memory. The models were created using Python and the PyTorch framework, leveraging its flexibility and comprehensive support for deep learning research. The training parameters were carefully selected to balance computational efficiency and model performance. All models trained for 10 epochs. To optimize training time and GPU memory usage, a batch size of 128 was chosen for training, while a batch size of 64 was used for validation and testing to balance memory consumption and processing efficiency.

The Adam optimizer was used with a learning rate of 0.0001, known for its ability to adjust the learning rate for each parameter, thus facilitating faster convergence. The models' weights were initialized using pre-trained weights from the ImageNet dataset to enable transfer learning. This approach utilizes features learned from a large dataset, enhancing the model's ability to quickly and efficiently converge on the specific task of classifying histopathological images. This configuration ensures a thorough and comprehensive assessment of the models, enabling the determination of the most effective architecture for classifying IDC in breast histopathology images.

% \noindent\textbf{Results}
\subsection{Results}

The ResNet-50 model illustrated improved accuracy and reduced loss during both the training and validation stages. The model achieved an overall accuracy rate of 92\%. For non-cancerous predictions, it had a precision of 0.94 and a recall of 0.95, while malignant predictions had a precision of 0.85 and a recall of 0.82. The F1-scores were 0.95 for non-cancerous and 0.84 for malignant classifications, demonstrating a balanced precision-recall tradeoff, crucial for accurate medical diagnosis.
Similarly, the DenseNet-121 model was achieving an overall accuracy of 92\%. For non-cancerous cases, it had a precision of 0.94 and a recall of 0.96, while for cancerous cases, it had a precision of 0.87 and a recall of 0.82. The F1-scores were 0.95 for non-cancerous and 0.84 for malignant classifications, indicating the model's high accuracy in categorizing histopathological images. 
The ResNeXt-50 model achieved an overall accuracy rate of 92\%. It had a precision of 0.94 and a recall of 0.95 for non-cancerous predictions, while malignant predictions had a precision of 0.86 and a recall of 0.83. The F1-scores were 0.95 for non-cancerous and 0.85 for cancerous classifications, indicating strong performance in classifying medical images.

The GoogLeNet (Inception v3) model also achieved an overall accuracy of 93\%. Non-cancerous predictions had a precision of 0.93 and a recall of 0.95, while malignant predictions had a precision of 0.86 and a recall of 0.89. The F1-scores were 0.95 for non-cancerous and 0.85 for malignant classifications, demonstrating a robust balance between precision and recall.
The EfficientNet model attained the highest accuracy among the CNN models at 93\%. Non-cancerous predictions had a precision of 0.96 and a recall of 0.94, while malignant predictions had a precision of 0.84 and a recall of 0.90. The F1-scores were 0.95 for non-cancerous and 0.87 for cancerous classifications, showing exceptional performance in medical image classification.
The MobileNetV2 model achieved an accuracy rate of 90\%. Non-cancerous predictions had a precision of 0.94 and a recall of 0.93, while malignant predictions had a precision of 0.82 and a recall of 0.85. The F1-scores were 0.93 for non-cancerous and 0.83 for cancerous classifications, demonstrating effectiveness with limited computational resources.
The SqueezeNet model attained an overall accuracy rate of 88\%. Non-cancerous predictions had a precision of 0.95 and a recall of 0.88, while malignant predictions had a precision of 0.74 and a recall of 0.87. The F1-scores were 0.91 for non-cancerous and 0.80 for cancerous classifications, indicating performance in a resource-limited setting.
The Vision Transformer (ViT) model demonstrated superior performance compared to the other models. It achieved an overall accuracy of 93\%. Non-cancerous predictions had a precision of 0.94 and a recall of 0.96, while malignant predictions had a precision of 0.89 and a recall of 0.84. The F1-scores were 0.95 for non-cancerous and 0.87 for malignant classifications, demonstrating excellent capacity to balance precision and recall.

Table \ref{Table2} presents a comprehensive overview of the performance outcomes of different deep learning models assessed in this research. It includes precise measurements such as precision, recall, F1-score, and total accuracy metrics for both non-IDC and IDC classifications. This thorough comparison evaluates the performance of the models, offering insights into their efficacy for classifying breast cancer histopathology images, and includes a ranking of their relative performance.

\begin{table*}[htbp]
  \centering
  \caption{Performance comparison of deep learning models for IDC classification in breast histopathology images} \label{Table2}
  \resizebox{0.85\textwidth}{!}{
  \begin{tabular}{l|c|cc|cc|cc|c}
    \toprule
    \multirow{2}{*}{Methods} & \multirow{2}{*}{Accuracy} & \multicolumn{2}{c}{Precision} & \multicolumn{2}{c}{Recall} & \multicolumn{2}{c}{F1-Score} & \multirow{2}{*}{Rank (std)} \\ 
    & & Class 0 & Class 1 & Class 0 & Class 1 & Class 0 & Class 1 &   \\ \midrule
    Resnet-50 & 0.92 & 0.94 & 0.86 & 0.95 & 0.82 & 0.95 & 0.84 & 4.57 (0.051) \\
    \rowcolor{gray!10} 
    DenseNet-121 & 0.92 & 0.94 & 0.87 & 0.96 & 0.82 & 0.95 & 0.84 &  4.64  (0.056) \\
    ResNeXt-50 & 0.92 & 0.94 & 0.86 & 0.95 & 0.83 & 0.95 & 0.85 & 5.14    (0.055) \\
    \rowcolor{gray!10} 
    Inception v3 & 0.92 & 0.94 & 0.85 & 0.95 & 0.85 & 0.96 & 0.85 & 4.57    (0.049) \\
    EfficientNet & 0.93 & 0.96 & 0.84 & 0.94 & 0.90 & 0.95 & 0.87 & 3.29    (0.044) \\
    \rowcolor{gray!10} 
    MobileNetV2 & 0.91 & 0.95 & 0.83 & 0.94 & 0.86 & 0.94 & 0.84 & 5.71    (0.051) \\
    SqueezeNet & 0.89 & 0.96 & 0.75 & 0.89 & 0.88 & 0.92 & 0.81 & 6.21 (0.070) \\
    ViT & 0.94 & 0.95 & 0.90 & 0.97 & 0.85 & 0.96 & 0.88 & 1.86 (0.045)  \\
    \bottomrule
  \end{tabular}
  }
\end{table*}

%%%%%%%%%%%%%%%%%%%
% \begin{table*}[htbp]
%   \centering
%   \caption{Performance comparison of deep learning models for IDC classification in breast histopathology images} \label{Table2}
%   \resizebox{0.85\textwidth}{!}{
%   \begin{tabular}{l|c|cc|cc|cc|c}
%     \toprule
%     \multirow{2}{*}{Methods} & \multirow{2}{*}{Accuracy} & \multicolumn{2}{c}{Precision} & \multicolumn{2}{c}{Recall} & \multicolumn{2}{c}{F1-Score} \\ 
%     & & Class 0 & Class 1 & Class 0 & Class 1 & Class 0 & Class 1 & Rank (std) \\ \midrule
%     Resnet-50 & 0.92 & 0.94 & 0.86 & 0.95 & 0.82 & 0.95 & 0.84 \\
%     \rowcolor{gray!10} 
%     DenseNet-121 & 0.92 & 0.94 & 0.87 & 0.96 & 0.82 & 0.95 & 0.84 \\
%     ResNeXt-50 & 0.92 & 0.94 & 0.86 & 0.95 & 0.83 & 0.95 & 0.85 \\
%     \rowcolor{gray!10} 
%     Inception v3 & 0.92 & 0.94 & 0.85 & 0.95 & 0.84 & 0.95 & 0.85 \\
%     EfficientNet & 0.93 & 0.96 & 0.84 & 0.94 & 0.90 & 0.95 & 0.87 \\
%     \rowcolor{gray!10} 
%     MobileNetV2 & 0.91 & 0.95 & 0.83 & 0.94 & 0.86 & 0.94 & 0.84 \\
%     SqueezeNet & 0.89 & 0.96 & 0.75 & 0.89 & 0.88 & 0.92 & 0.81 \\
%     ViT & 0.94 & 0.95 & 0.90 & 0.97 & 0.85 & 0.96 & 0.88 \\
%     \bottomrule
%   \end{tabular}
%   }
% \end{table*}
%%%%%%%%%%%%%%%%%%%

% \noindent\textbf{Comparison and Analysis}
\subsection{Comparison and Analysis}

The ViT model showed improved performance compared to conventional CNN designs, achieving a superior overall accuracy of 93\%, surpassing other models with accuracy rates ranging from 88\% to 92\%. The precision and recall values of ViT for non-cancerous (0.94 and 0.96, respectively) and malignant (0.89 and 0.84, respectively) predictions demonstrate a high degree of reliability in accurately distinguishing between the two classes. The F1-scores of 0.95 for non-cancerous and 0.87 for cancerous classifications highlight the system's strong ability to properly handle both positive and negative samples.

\begin{figure*}
\centering
\includegraphics[width=0.6\linewidth]{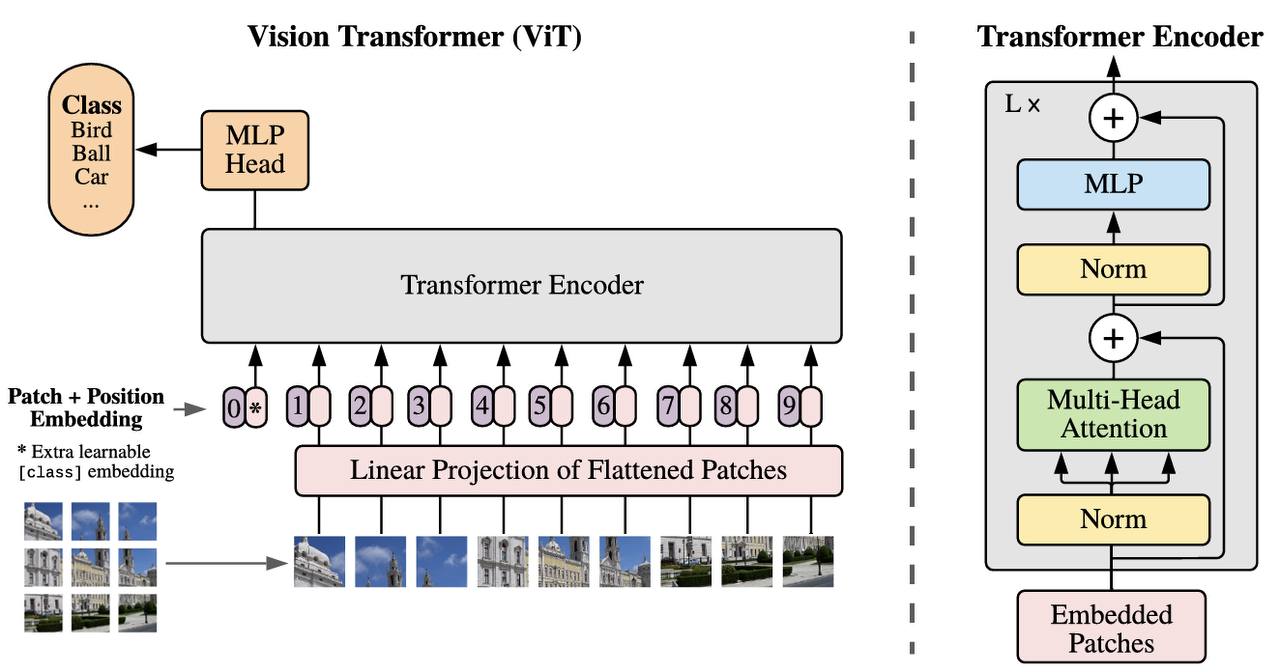}
\caption{Vision Transformer model architecture \cite{ref17-xie2017aggregated}}\label{fig:ViT}
\end{figure*}

EfficientNet, one of the CNN models, achieved a maximum accuracy of 92\%, demonstrating excellent precision and recall metrics. However, it did not quite match the performance level demonstrated by the ViT model. ResNet, DenseNet, ResNeXt, and GoogLeNet all attained a 91\% accuracy, showcasing their efficacy, albeit with somewhat inferior precision and recall compared to EfficientNet and ViT. MobileNetV2 and SqueezeNet, although computationally efficient, exhibited lower accuracy (90\% and 88\%, respectively) and somewhat diminished performance metrics, making them less effective than the other models for this particular task.

The ViT model is the most successful architecture for classifying invasive ductal carcinoma (IDC) in breast histopathology images. It achieves maximum accuracy and maintains a well-balanced precision, recall, and F1-score. The utilization of self-attention mechanisms in capturing both global and local information within images provides it with a distinct advantage over conventional CNN architectures.

\section{CONCLUSION AND FUTURE WORK}

This study conducted a thorough comparative investigation of various deep learning models for classifying histopathology images of breast cancer, specifically differentiating between Invasive Ductal Carcinoma (IDC) and non-IDC tissue samples. The models assessed included ResNet-50, DenseNet-121, ResNeXt-50, GoogLeNet (Inception v3), EfficientNet, MobileNetV2, SqueezeNet, and Vision Transformer (ViT).

The experimental findings revealed that the ViT model outperformed conventional CNN architectures in terms of overall accuracy, precision, recall, and F1-score. The ViT model achieved a remarkable accuracy rate of 93\%, with F1-scores of 0.95 for non-IDC and 0.87 for IDC, demonstrating its ability to maintain a balance between precision and recall. EfficientNet showed the highest accuracy among the CNN models at 92\%, followed closely by ResNet, DenseNet, ResNeXt, and GoogLeNet, each achieving 91\% accuracy. MobileNetV2 and SqueezeNet, although computationally efficient, showed lower accuracy levels of 90\% and 88\%, respectively.
The exceptional performance of the ViT model highlights the capacity of attention-based mechanisms to capture both comprehensive and specific characteristics essential for precise medical image interpretation. This work underscores the importance of using advanced deep learning architectures to enhance diagnostic accuracy in clinical settings, leading to better patient outcomes.

While this study provides valuable insights, future research should explore advanced data augmentation techniques and preprocessing methods to improve model performance, particularly in addressing class imbalance and enhancing generalization. Additionally, a model based on Mamba will be explored for breast cancer diagnosis with mammography images to improve accuracy and efficiency \cite{ref9-nasiri2024vision}. Investigating ensemble approaches and developing real-time implementations of these models in clinical settings are also important areas for further research to assess practical utility and integration into diagnostic workflows.

Improving the interpretability of deep learning models is crucial for their acceptance and use in clinical practice. Future research should focus on developing methods to provide a deeper understanding of model decisions, enhancing transparency for medical practitioners. Exploring transfer learning and domain adaptation strategies could expand the applicability of these models to other types of cancer or medical imaging tasks. Using larger and more diverse datasets will help evaluate the models' robustness and generalizability across different populations and imaging conditions. Integrating histopathological image analysis with additional clinical data, such as patient demographics and genetic information, has the potential to improve diagnostic accuracy and provide a more comprehensive understanding of the disease.

In summary, this study demonstrates that advanced deep learning models, particularly the Vision Transformer, can significantly improve the classification of breast cancer from histopathology images. Further exploration of this field, with a focus on the proposed future directions, holds promise for advancing medical image analysis and improving clinical outcomes.

\newpage

\bibliographystyle{unsrt}

% \bibliography{report} 
% \bibliographystyle{spiebib} 

% \bibliographystyle{unsrt}
% \bibliography{report.bib}

\end{document}